\documentclass[aps,prc,amsmath,showpacs,showkeys,nofootinbib,
superscriptaddress]{revtex4}
\usepackage[dvips]{graphicx}

\begin{document} 

\title{\bf $\Lambda$-hypernuclear production in ($K^-_{\rm stop}, \pi$) 
reactions reexamined}

\author{V.~Krej\v{c}i\v{r}\'{i}k} 
\email{v.krejcirik@ujf.cas.cz} 
\affiliation{Faculty of Mathematics and Physics, 
Charles University, 12116 Prague, Czech Republic} 
\affiliation{Nuclear Physics Institute, 25068 \v{R}e\v{z}, Czech Republic} 

\author{A.~Ciepl\'{y}} 
\email{cieply@ujf.cas.cz} 
\affiliation{Nuclear Physics Institute, 25068 \v{R}e\v{z}, Czech Republic} 

\author{A.~Gal} 
\email{avragal@vms.huji.ac.il} 
\affiliation{Racah Institute of Physics, The Hebrew University, 
91904 Jerusalem, Israel} 

\date{\today} 
  
\begin{abstract}

Distorted wave impulse approximation calculations of $\Lambda$-hypernuclear 
production rates in stopped $K^-$ reactions on several $p$-shell targets used 
recently in experiments by the FINUDA Collaboration are reported. Chirally 
motivated $K^- N \to \pi\Lambda$ in-medium transition amplitudes are employed 
and the sensitivity of the calculated rates to the initial $K^-$-atomic 
wave functions and final pion distorted waves is studied. The calculated rates 
are compared with measured rates, wherever available, confirming earlier 
observations that (i) the calculated rates are generally lower than the 
measured rates, and (ii) the deeper the $K^-$-nuclear potential, the worse is 
the discrepancy. The $A$ dependence of the calculated $1s_{\Lambda}$ 
production rates is discussed for the first time, providing a useful tool to 
resolve the issue of depth of the $K^-$-nuclear potential near threshold. 

\end{abstract}

%21.80.+a: Hypernuclei 
%21.85.+d: Mesic nuclei 
%25.80.Nv: Kaon-induced reactions 
%36.10.Gv: Mesonic atoms 


\maketitle 

%\newpage 

\section{Introduction}

$\Lambda$-hypernuclear production in $(K^-_{\rm stop},\pi)$ reactions, 
in which the final state is uniquely identified by measuring the outgoing 
pion momentum, was reported for the first time in stopped $K^-$ experiments 
at CERN in 1973 \cite{Faessler} and more recently in experiments at 
KEK \cite{Tamura}, BNL \cite{Ahmed} and at DA$\Phi$NE, Frascati, by 
the FINUDA Collaboration \cite{Agnello,HYP06,BonomiHYPX}. On the theoretical 
side, several distorted wave impulse approximation (DWIA) calculations 
of $(K^-_{\rm stop},\pi)$ hypernuclear production rates have been 
reported~\cite{HLW74,GalKlieb,MatsuyamaYazaki,CieplyFriedman,KC10}, but none 
of them led to satisfactory agreement with the measured rates.{\footnote
{Ref.~\cite{KC10} is a preliminary conference version which is outdated by 
the present paper.}} In general, these calculated capture rates fall below the 
experimentally reported rates, with the exception of the old CERN data for 
$^{12}{\rm C}$ \cite{Faessler}. 

The present paper primarily covers the production of $^{7}_{\Lambda}{\rm Li}$, 
$^{9}_{\Lambda}{\rm Be}$, $^{12}_{~\Lambda}{\rm C}$, $^{13}_{~\Lambda}{\rm C}$ 
and $^{16}_{~\Lambda}{\rm O}$, for all of which preliminary data have recently 
been reported \cite{BonomiHYPX}. We focus on the $A$ dependence of the 
calculated rates, hitherto not explored systematically, to look for further 
tests of the role played by initial- and final-state interactions. In 
conjunction with previous calculations, we use the DWIA. Several $K^-$ nuclear 
optical potentials are used to generate the required initial-state $K^-$-atom 
wave functions, and several pion nuclear optical potentials are used to 
generate final-state pion distorted waves (DWs). 
The underlying $K^- N \to \pi \Lambda$ reaction amplitude is studied in 
free space, as well as in the nuclear medium, within the chiral Lagrangian 
framework \cite{CieplyFriedman,KaiserSiegelWeise,WaasKaiserWeise,OsetRamos,
BorasoyNisslerWeise,CieplySmejkal,CieplySmejkalPrep,Lutz} to generate 
in-medium branching ratios (BRs) for stopped $K^-$ reactions. Past 
works \cite{GalKlieb,MatsuyamaYazaki,CieplyFriedman} used BRs extrapolated 
from emulsion experiments \cite{VanDerVelde}. We compare results obtained in 
both approaches. 

The present paper is organized as follows. 
The capture at rest DWIA formalism is outlined in Sec.~II. The choice of the 
microscopic chiral model for $K^- N \to \pi Y$ reactions at rest, together 
with the BRs derived in this model and the input wave functions to the DWIA 
calculations, are specified in Sec.~III. Results of $\Lambda$ hypernuclear 
production rate calculations for stopped $K^-$ reactions are presented and 
discussed in Sec.~IV, with a brief conclusion given in Sec.~V.

\section{Capture at rest calculations}

We study the reaction 
\begin{equation} 
K^{-}\;+\;A(i)\;\longrightarrow \;\pi^{-\tau-1/2}\;+\;H(f)
\label{reakcecela} 
\end{equation} 
in which a $K^{-}$ meson is captured on a target nucleus denoted as $A$ in 
its ground state $i$, from an atomic orbit $nL$ into a final state $f$ of 
a $\Lambda$ hypernucleus $H$ plus an outgoing pion. The superscript 
$-\tau-1/2$ denotes the pion charge ($\tau=\pm 1/2$ for $\pi^-$ and $\pi^0$ 
respectively). We follow the capture at rest calculation formalism detailed in 
Ref.~\cite{GalKlieb}. In the DWIA, the nuclear reaction Eq.~(\ref{reakcecela}) 
is induced by the one-baryon transition 
\begin{equation} 
K^{-}\;+\;N\;\longrightarrow \;\pi^{-\tau-1/2}\;+\;Y 
\label{reakce1nukl} 
\end{equation} 
on a nucleon $N$ to a hyperon $Y$, with the in-medium $T$ matrix assumed here 
to be of $s$-wave type: 
\begin{equation} 
T_{fi}({\bf q}_{f})= 
  \sum_{j=1}^{\cal N} \:\langle \,f\mid\:t_{j}({\bf q}_{f})\:\mid i\, \rangle 
  = t({q}_{f})\,\rho_{fi}^{DW}({\bf q}_{f}). 
\label{eq:T} 
\end{equation} 
The charge indices are omitted for simplicity and the DW transition 
form factor is given by 
\begin{equation} 
\rho_{fi}^{DW}({\bf q}_{f})=\int d^{3}r\: \chi^{(-)*}_{{\bf q}_{f}}({\bf r}) 
  \:\rho_{fi}({\bf r})\:\Psi_{nLM}({\bf r}), 
\label{eq:DW} 
\end{equation} 
where $\rho_{fi}$ stands for the nuclear to hypernuclear transition 
matrix element. The $K^{-}$-atomic wave function $\Psi_{nLM}$ is obtained by 
solving the Klein-Gordon equation with a $K^{-}$-nuclear strong interaction 
optical potential $V^{K}_{\rm opt}$ added to the Coulomb potential $V_C$ 
generated by the nuclear charge distribution plus vacuum polarization. The 
dependence on the magnetic quantum number $M$ was suppressed on the left-hand 
side (l.h.s.) of Eq.~(\ref{eq:DW}). The pion DW $\chi^{(-)}_{{\bf q}_{f}}$ in 
the final state is given in terms of a partial-wave expansion: 
\begin{equation} 
\chi^{(-)*}_{{\bf q}_{f}}({\bf r})=\sum_{\ell} (-{\rm i})^{\ell}(2{\ell}+1) 
  \tilde{j}_{\ell}(r)\,P_{\ell}(\hat{\bf q}\!\cdot\!\hat{\bf r}). 
\end{equation} 
The radial wave function $\tilde{j}_{\ell}(r)$, which reduces to the spherical 
Bessel function $j_{\ell}(qr)$ for a free pion, solves the Klein-Gordon 
equation with the pion-nuclear optical potential, plus the appropriate 
electromagnetic potential for a charged pion. 

The nuclear capture rate per stopped $K^{-}$ in the reaction 
Eq.~(\ref{reakcecela}) is given by 
\begin{equation} 
R_{fi}/K^{-}= \frac{q_f\omega_f}{{\bar q}_f{\bar\omega}_f} 
  \:R(K^-N\rightarrow\pi\Lambda)\: 
  \frac{\int d\Omega_{q_f} \langle\,\mid\rho_{fi}^{DW}({\bf q}_f) 
        \mid^{2}\,\rangle} 
       {4\pi {\bar\rho}_N}, 
\label{eq:R/K} 
\end{equation} 
where the fractions $R(K^-N\rightarrow\pi\Lambda)$ are the elementary BRs for 
mononucleonic $K^{-}$ absorption at rest in the nuclear medium. The brackets 
$\langle \cdots \rangle$ mean that the absolute square of the DW transition 
form factor is to be averaged on the initial states and summed over the final 
ones. The kinematical factor in front of $R$ in Eq.~(\ref{eq:R/K}) appears 
because of transformation of the two-body scattering amplitude, which 
describes the elementary reaction Eq.~(\ref{reakce1nukl}), into the many-body 
center-of-mass frame. The momentum $q_{f}$ of the outgoing pion is determined 
by energy conservation, and $\omega_{f}$ stands for the reduced energy in the 
final state, 
\begin{equation} 
\omega_{f}^{-1}=E_{\pi}^{-1}(q_{f})+E_{H}^{-1}(q_{f})\;\; ;\;\; 
\omega_{f}\longrightarrow E_{\pi}(q_{f})\;\;\;\mbox{for }A\rightarrow \infty, 
\end{equation} 
where ${\bar q}_f$, ${\bar\omega}_f$ are appropriately averaged 
in-medium quantities. Finally, ${\bar\rho}_N$ denotes the effective 
nuclear density available to the $K^-$ capture process, 
\begin{equation} 
{\bar\rho}_N = \int \rho_{N}(r) \mid R_{nL}(r)\mid^{2} \: r^2 \: dr, 
\label{efektivnihustota} 
\end{equation} 
where the nucleon density $\rho_{N}$ and the $K^-$-atomic radial 
wave function are normalized according to  
\begin{equation} 
\int \rho_{N}\:d^{3}r = {\cal N}, \;\;\;\;\;\; 
\int \mid R_{nL}(r)\mid^{2} \: r^2 \: dr = 1, 
\label{norm} 
\end{equation} 
where ${\cal N}$ denotes the number of neutrons or protons for 
$\tau = \pm 1/2$, respectively. 

The last factor on the right-hand side (r.h.s.) of Eq.~(\ref{eq:R/K}), 
\begin{equation} 
R_{fi}/Y=\frac{\int d\Omega_{{q}_f} \langle\,\mid\rho_{fi}^{DW}
       ({\bf q}_f)\mid^{2}\,\rangle} 
       {4\pi {\bar\rho}_N}, 
\end{equation} 
is loosely termed the capture rate per hyperon $Y$ because its derivation 
assumes that the capture reaction Eq.~(\ref{reakce1nukl}) is the only one 
available. It can be decomposed into contributions, which correspond to 
transitions with multipolarity $k$ from a given $n_{N}l_{N}$ nuclear shell to 
a given $n_{Y}l_{Y}$ hypernuclear shell, in the following form \cite{GalKlieb}: 
\begin{equation} 
R^{k}_{n_{N}l_{N}\rightarrow n_{Y}l_{Y}}= {\cal N}(n_{N}l_{N}) 
  \frac{(2k+1)\,(l_{N}\: 0\: k\: 0 \mid l_{Y}\: 0)^{2} 
         N^{k}_{n_{Y}l_{Y},n_{N}l_{N}}} 
       {4\pi {\bar\rho}_N}. 
\label{eq:R/Y} 
\end{equation} 
Here, ${\cal N}(n_{N}l_{N})$ is the neutron (proton) occupation number 
of the target nuclear shell for $\tau=+1/2~(-1/2)$, the Clebsch-Gordan 
coefficient squared accounts for the conservation of angular momentum 
and parity, and the entities
\begin{equation} 
N^{k}_{n_{Y}l_{Y},n_{N}l_{N}}= \sum_{\ell} (L\: 0\: k\: 0 \mid {\ell}\: 0)^{2} 
 \mid I^{\ell}_{n_{Y}l_{Y},n_{N}l_{N}} \mid^{2} 
\label{eq:Nk} 
\end{equation} 
are the appropriate averages of the absolute squares of the DWIA amplitudes 
\begin{equation} 
I^{\ell}_{n_{Y}l_{Y},n_{N}l_{N}}= \int_{0}^{\infty} dr\: \tilde{j}_{\ell}(r)\: 
  u^{*}_{n_{Y}l_{Y}}(r)\: u_{n_{N}l_{N}}(r)\: R_{nL}(r), 
\label{eq:Il} 
\end{equation} 
where $u_{n_{B}l_{B}}(r)/r$ are the radial parts of the one-baryon wave 
functions. Eqs.~(\ref{eq:R/Y})-(\ref{eq:Il}) assume that the DWIA capture rate 
calculation does not depend on the total angular momenta $j_{B}=l_{B}\pm 1/2$ 
for the orbits in question. This was justified by the numerical calculation 
performed in Ref.~\cite{GalKlieb}, where more general formulae for the 
dependence on $j_{B}$ can be found as well, and is checked later in the 
present paper.

\section{Input} 

In this section, we specify the entities that are needed to perform numerical 
calculations of nuclear capture rates. First, we outline the model used for 
the one-baryon capture process Eq.~(\ref{reakce1nukl}). Subsequently, we 
specify the baryon and meson nuclear potentials chosen to generate initial- 
and final-state wave functions for use in the evaluation of the DWIA 
amplitudes Eq.~(\ref{eq:Il}) that serve as input to nuclear capture rate 
calculations. 

\subsection{$K^-N\rightarrow\pi\Lambda$ branching ratios}

\begin{figure} 
\includegraphics[scale=0.4]{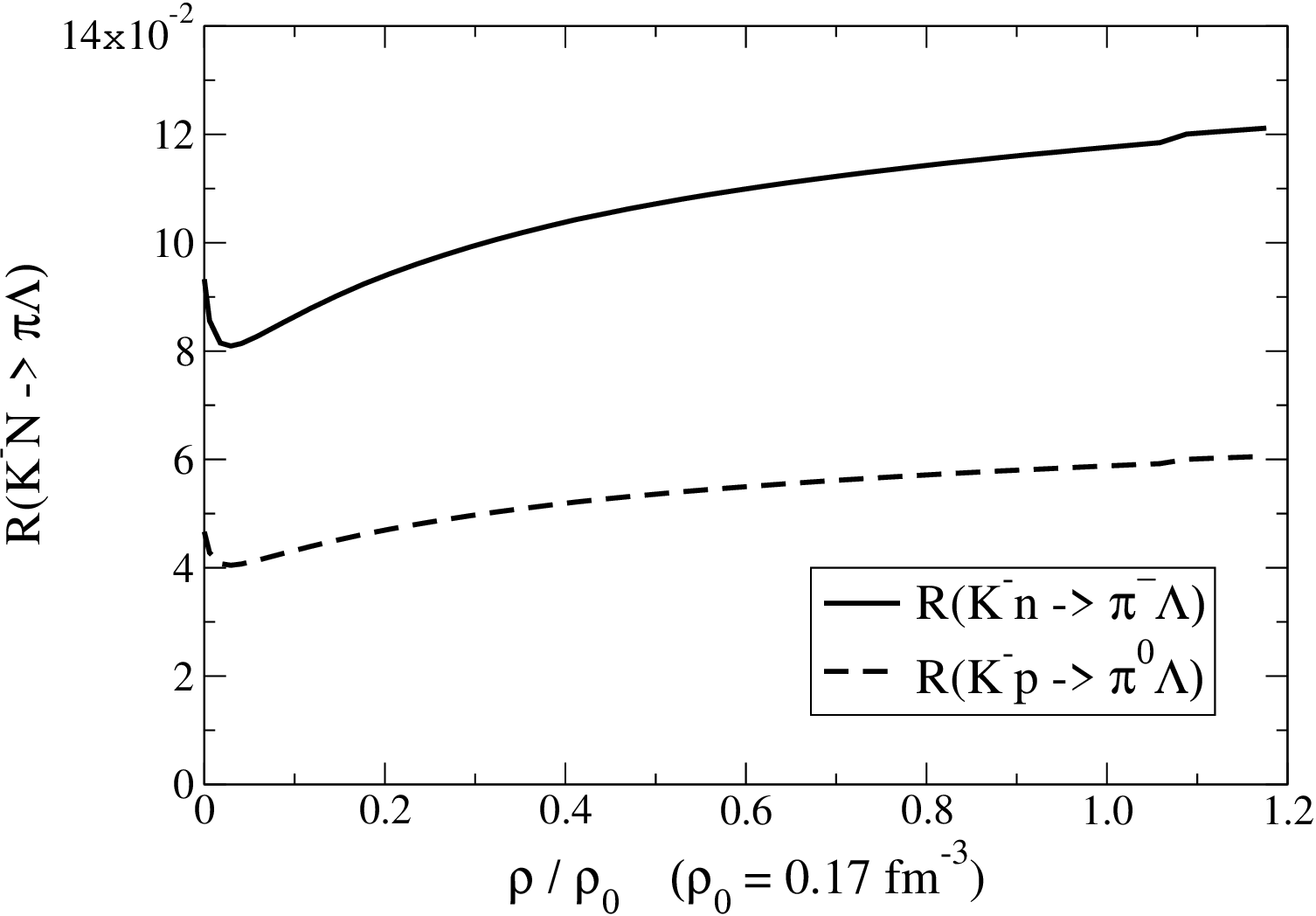} 
\caption{Calculated branching ratios $R(K^-N\to\pi\Lambda)$ 
as a function of the nuclear density for capture on neutrons (solid curve) 
and on protons (dashed curve).} 
\label{figBRratios} 
\end{figure} 

For the calculation of BRs $R(K^-N\rightarrow\pi\Lambda)$ 
of the one-baryon capture process Eq.~(\ref{reakce1nukl}), we adopted the 
effective potential model based on chiral symmetry, as described in detail in 
Refs.~\cite{KaiserSiegelWeise,WaasKaiserWeise,CieplySmejkal,CieplySmejkalPrep,
Lutz}. The required BRs are obtained in coupled-channel calculations that 
include ten meson-baryon channels coupled to the $K^-p$ system 
\cite{CieplySmejkal,CieplySmejkalPrep}. We also take the effects of Pauli 
blocking and kaon self-energy in the nuclear medium into account \cite
{CieplyFriedman,WaasKaiserWeise,Lutz}. The dependence of the calculated 
BRs on the nuclear density is demonstrated in Fig.~\ref{figBRratios} 
as a function of the fractional nuclear density $\rho/\rho_0$, where 
$\rho_0=0.17~{\rm fm}^{-3}$ is nuclear-matter density. Although the central 
nuclear density varies along the periodic table roughly in the range $0.14-
0.22~{\rm fm}^{-3}$, the BRs shown in the figure do not change much over 
this range of densities. Therefore, the precise dependence on $\rho$ may be 
neglected, and we assume that the $K^-N\rightarrow\pi\Lambda$ capture 
reaction takes place at a proton (or neutron) density $\rho=\rho_0/2$. 
For further applications, we denote the BRs obtained at nuclear density 
$\rho=\rho_0/2$ by BR1, and the BRs obtained in vacuum by BR2. 

The use of a microscopic model for the 
$K^-N\rightarrow\pi\Lambda$ BRs is one of the novelties of the present paper. 
Past works used BRs derived indirectly by extrapolating from measurements 
performed in carbon and freon emulsions \cite{VanDerVelde}, a procedure that 
is prone to systematic errors. We use these emulsion BRs (labeled here as BR3) 
to compare with BRs obtained from our microscopic chiral model. This is shown 
in Table~\ref{TabBRratios}, where the maximum difference between the various 
BRs (BR1, BR2, and BR3) is as large as $30\%$. However, this variation is 
still small compared to other effects discussed below. The ratio of 1/2 for 
BRs on a proton to BRs on a neutron follows from charge independence, which is 
implemented by conserving isospin in our model for $K^-N\rightarrow\pi\Lambda$. 

\begin{table} 
\caption{$R(K^-N\to\pi\Lambda)$ branching ratios (in units of $10^{-2}$).} 
\label{TabBRratios} 
\begin{ruledtabular} 
\begin{tabular}{lclcclccl} 
& branching ratio &&BR1&BR2&& \multicolumn{2}{c}{BR3~\cite{VanDerVelde}} & \\
& $R(K^-N\rightarrow\pi\Lambda)$ && $\rho=\rho_0/2$ & $\rho=0$ && 
$^{12}{\rm C}$ & $^{16}{\rm O}$ & \\  \hline 
& $R(K^-n \rightarrow \pi^- \Lambda)$ && 10.54 & 9.68 && 8.7 & 7.7 & \\  
& $R(K^-p \rightarrow \pi^0 \Lambda)$ && 5.27 & 4.84 && 4.4 & 3.9 & \\  
\end{tabular} 
\end{ruledtabular} 
\end{table} 

\subsection{Wave functions}

To perform the numerical calculation of the DWIA integrals 
$I^{\ell}_{n_{Y}l_{Y},n_{N}l_{N}}$ Eq.~(\ref{eq:Il}) and 
${\bar \rho}_N$ Eq.~(\ref{efektivnihustota}), wave functions of the 
initial kaon and nucleon and of the final hyperon and outgoing pion are 
needed. These wave functions are generated by solving the respective wave 
equations with fitted potentials. 

\subsubsection{$K^-$ wave functions} 

For the kaon wave function, we use the Klein-Gordon equation with a potential 
which consists of two parts, the finite-size Coulomb potential plus 1st-order 
vacuum polarization corrections, and the strong-interaction optical potential 
parametrized phenomenologically by the form \cite{Batty}: 
\begin{equation} 
V^{K}_{\rm opt}(r) = - \frac{4 \pi}{2 \mu_K} \left(1+\frac{\mu_K}{M_N}\right) 
\left[b+B \left(\frac{\rho(r)}{\rho(0)} \right)^{\nu} \, \right] \rho(r). 
\label{optpot} 
\end{equation} 
Here, $\mu_K$ stands for the kaon-nucleus reduced mass, $M_N$ is the nucleon 
mass and $\rho(r)$ denotes the nuclear density normalized to the number of 
nucleons $A$. We use three different parameter sets for the kaon-nucleus 
optical potential, as specified in Table~\ref{TabKaonOptPot}. In the last 
column of the table, for orientation, we show the approximate depth of the 
strong-interaction (real part) potentials at nuclear density 
$\rho=\rho_0=0.17~{\rm fm}^{-3}$. 

\begin{table} 
\caption{Parameters of the $K^-$ nuclear optical potential Eq.~(\ref{optpot}).} 
\label{TabKaonOptPot} 
\begin{ruledtabular} 
\begin{tabular}{llccccl} 
& potential & $b~[{\rm fm}]$ & $B~[{\rm fm}]$ & $\nu$ & 
Re~$V^{K}_{\rm opt}(\rho=\rho_0)$\, [MeV] & \\ \hline 
& $[K_{\chi}]$ & $0.38+0.48{\rm i}$ & $0$ & $0$ &   $-50$ &  \\  
& $[K_{\rm eff}]$ & $0.63+0.89{\rm i}$ & $0$ & $0$ & $-80$ & \\ 
& $[K_{\rm DD}]$ & $-0.15+0.62{\rm i}$ & $1.65-0.06{\rm i}$ & $0.23$ & 
$-190$ & \\ 
\end{tabular} 
\end{ruledtabular} 
\end{table} 

The meson-nuclear optical potential is often expressed by an effective 
scattering length multiplied by the nuclear density. Thus, the parameter 
$b$ for the potential $[K_{\chi}]$ represents the average of the $K^-n$ 
and $K^-p$ scattering lengths in the nuclear medium computed using the chiral 
model discussed in Sec.~III A. The values of parameter sets for 
potentials $[K_{\rm eff}]$ and $[K_{\rm DD}]$ were fitted to reproduce a large 
set of kaonic atom data by Friedman {\it et al.} \cite{Batty}. For $B=0$, the 
potential reduces to the standard `effective' $[K_{\rm eff}]$ parametrization 
of the optical potential. The potential $[K_{\rm DD}]$ explicitly exhibits 
a substantial density dependence, which may be related to the dynamics of the 
$\Lambda(1405)$ resonance submerged into the nuclear medium \cite{Batty}. 

For the targets considered in the present paper, the relevant $K^-$-atomic 
orbits are represented by the $2P$ ($L=1$) and $3D$ ($L=2$) states. 
These are the lowest $L$ orbits observed in the X-ray cascade and are 
sufficiently close to the nucleus for capture to occur significantly. 
The calculations were done separately for each of these orbits, and a weighted 
average was then taken according to Batty's analysis of the atomic cascade 
process \cite{BattyComunication}. The relative populations of these $L = 1,2$ 
atomic orbits (summed over $n$) are listed in Table~\ref{TabRelPop}. 

\begin{table} 
\caption{Relative population of $K^-$-atomic orbits \cite{BattyComunication}.} 
\label{TabRelPop} 
\begin{ruledtabular} 
\begin{tabular}{lcccccl} 
& orbit & $^{7}{\rm Li}$ & $^{9}{\rm Be}$ & $^{12,13}{\rm C}$ & $^{16}{\rm O}$ 
& \\  \hline 
& $P$ &   0.76 & 0.49 & 0.23      & 0.18 & \\  
& $D$ &   0.24& 0.51  & 0.77      & 0.82 & \\  
\end{tabular} 
\end{ruledtabular} 
\end{table} 

\subsubsection{Baryon wave functions} 

The wave functions of nucleons and hyperons were computed numerically as bound 
states in a Woods-Saxon potential, 
\begin{equation} 
V(r) =-\frac{V_0}{1+\exp{(r-R)/a}}\,\,\,,\,\,\, R=r_0\,A^{1/3} \,, 
\label{WSpotential} 
\end{equation} 
with geometry fixed by setting $a=0.6\,\,{\rm fm}$ and $r_0=1.25\,\,{\rm fm}$. 
The potential depth $V_0$ was adjusted separately for each baryon state so 
that the corresponding binding energy was reproduced; 
see Ref.~\cite{HashimotoTamura} for a compilation of $\Lambda$ hypernuclear 
binding energies. 

\subsubsection{Pion distorted waves} 

The pionic optical potential is taken to be of the standard form 
\cite{EricsonEricson}, often used in the analysis of pionic atoms and 
pion-nuclear scattering: 
\begin{eqnarray} 
-\frac{2 \mu_{\pi}}{4 \pi} V^{\pi}_{\rm opt} &=& \left(1+\frac{m_{\pi}}{M_N} 
\right) b_0 \rho(r) + \left(1+\frac{m_{\pi}}{2M_N} \right)B_0 \rho^2(r) - 
{\bf \nabla}\frac{\alpha(r)}{1+\frac{4\pi}{3}\xi\alpha(r)} {\bf \nabla} \\ 
\alpha(r) &=& \left( 1+\frac{m_{\pi}}{M_N} \right)^{-1} c_0 \rho(r) + 
\left(1 + \frac{m_{\pi}}{2M_N} \right)^{-1} C_0 \rho^2(r). \nonumber 
\label{eq:piopt}
\end{eqnarray} 
Here, $\mu_{\pi}$ stands for the pion-nuclear reduced mass, $m_{\pi}$ and 
$M_N$ are the pion and nucleon masses, and $\rho(r)$ denotes the nuclear 
density normalized to the number of nucleons $A$. Calculations were performed 
with a free (plane-wave) pion and with two different parameter sets for the 
pion-nuclear optical potential. These potentials were fitted to low-energy 
scattering data \cite{pionoptpot1,pionoptpot2}, and their parameters are 
listed in Table~\ref{TabPionOptPot}. 

\begin{table}  
\caption{Parameters of the pionic optical potential Eq.~(\ref{eq:piopt}).} 
\label{TabPionOptPot}  
\begin{ruledtabular} 
\begin{tabular}{lccccccl} 
& potential & $b_0~[m_{\pi}^{-1}]$ & $B_0~[m_{\pi}^{-4}]$&$c_0~[m_{\pi}^{-3}]$ 
& $C_0~[m_{\pi}^{-6}]$ & $\xi$ & \\ \hline 
& $\pi_b$ & $0.268+0 {\rm i}$ & $0$ & $0.036+0.206{\rm i}$ & $0- 0.203 
{\rm i}$ & 1.4 & \\  
& $\pi_c$ & $0.010+0.437 {\rm i}$ & $0$ & $0.047+0.222{\rm i}$ &$0$&0& \\ 
\end{tabular} 
\end{ruledtabular} 
\end{table}

\section{Results and discussion}

Here, we present the results of calculations of $K^-$ capture rates for 
$\Lambda$ hypernuclear production on $p$-shell targets and discuss their 
sensitivity to various inputs. It was assumed in these calculations that 
capture to the low-lying $\Lambda$ hypernuclear bound states, or resonances, 
occurs only through baryon transitions $1p_N\rightarrow 1s_{\Lambda}$ and 
$1p_N\rightarrow 1p_{\Lambda}$ from the $1p_N$ valent shell to the 
$1s_{\Lambda}$ and $1p_{\Lambda}$ single-particle (s.p.) states, respectively, 
in the final hypernucleus. Since the experimental data on light hypernuclei 
often  do not indicate distinct hypernuclear s.p. structure for the 
$1p_{\Lambda}$ configuration, we considered the production of $1p_{\Lambda}$ 
states only beginning with $A=12$. Furthermore, we note that the 
multipolarity $k$ in Eqs.~(\ref{eq:R/Y}) and (\ref{eq:Nk}) is limited 
to $k=1$ ($1^-$ transition) for the $1p_N \rightarrow 1s_{\Lambda}$ 
capture process, whereas two values $k=0,2$ are allowed in the 
$1p_N \rightarrow 1p_{\Lambda}$ capture process ($0^+,2^+$ transitions). 
The number of valence nucleons ${\cal N}(1p_N)$ which contribute to the 
capture rates Eq.~(\ref{eq:R/Y}) is $2,~3,~4,~5,~6$ (neutrons) in the 
$(K^-_{\rm stop}, \pi^-)$ reaction for target nuclei $^{7}{\rm Li}$, 
$^{9}{\rm Be}$, $^{12}{\rm C}$, $^{13}{\rm C}$, $^{16}{\rm O}$, respectively. 

\subsection{Sensitivity tests} 

\subsubsection{Baryon wave functions} 

To test the sensitivity of our results to the baryon wave functions generated 
by Woods-Saxon potentials Eq.~(\ref{WSpotential}), we calculated the 
capture rates for the production of $^{12}_{~\Lambda}{\rm C}$ (in both the 
$1s_{\Lambda}$ and $1p_{\Lambda}$ states) for a modified geometry of the 
Woods-Saxon potential. Specifically, we used $A=11,13$ instead of $A=12$ in 
Eq.~(\ref{WSpotential}). The difference in capture rates was less than 10\%. 
We also varied the depth of the Woods-Saxon potential and checked that its 
variation by about 10\% leads to a less than 5\% difference in the capture 
rates. In general, realistic variations of the baryon wave functions have 
a relatively small impact (less than 10\%) on the calculated capture rates. 

A further sensitivity test is demonstrated here for $^{16}{\rm O}$ target. 
Table~\ref{TabNuclSensO} lists capture rates per hyperon [Eq.~(\ref{eq:R/Y}) 
per one $p$-shell neutron, ${\cal N}(n_{N}l_{N})=1$], obtained for a neutron 
in each one of the nuclear $1p_j$ subshells (with binding energies which 
differ roughly by 6 MeV), for a $1s_{\Lambda}$ hyperon with binding energy 
given by the $\Lambda$ separation energy in the hypernuclear ground state and 
for each one of the $2P$ and $3D$ $K^-$ orbits. Except for the case of $2P$ 
and [$K_{\chi}$], switching from $1p_{1/2}$ to $1p_{3/2}$ makes little 
difference, in agreement with the assumption that the relevant entities depend 
only weakly on the total angular momenta $j_B=l_B \pm 1/2$. 
However, the $2P$ orbit contribution is considerably weaker than that of the 
$3D$ orbit contribution in $^{16}{\rm O}$ (see also Table~\ref{TabRelPop}). 
Hence, this $j$ dependence is negligible for the $K^-$ [$K_{\chi}$] potential, 
whereas it amounts to approximately a $10\%$ effect for the $K^-$ 
[$K_{\rm DD}$] potential. 

\begin{table} 
\caption{Capture rates per hyperon and per one $p$-shell neutron (in units 
of $10^{-4}$) calculated for different $j$ orbits assumed for $1p$ shell 
neutrons in $^{16}{\rm O}$, see text.} 
\label{TabNuclSensO} 
\begin{ruledtabular} 
\begin{tabular}{lcccccl} 
& \multicolumn{1}{c}{$K^-$ orbit} & \multicolumn{2}{c}{[$K_{\chi}$]} & 
\multicolumn{2}{c}{[$K_{\rm DD}$]} & \\
&  & $(1p_{1/2})_n$ & $(1p_{3/2})_n$ & 
$(1p_{1/2})_n$ & $(1p_{3/2})_n$ & \\ \hline
& $2P$ & 1.23 & 1.43 & 0.33 & 0.37 & \\  
& $3D$ & 2.64 & 2.67 & 0.66 & 0.72 & \\ 
\end{tabular} 
\end{ruledtabular} 
\end{table} 

\subsubsection{$K^-$ capture branching ratios and wave functions} 

\begin{figure} 
\includegraphics[scale=0.4]{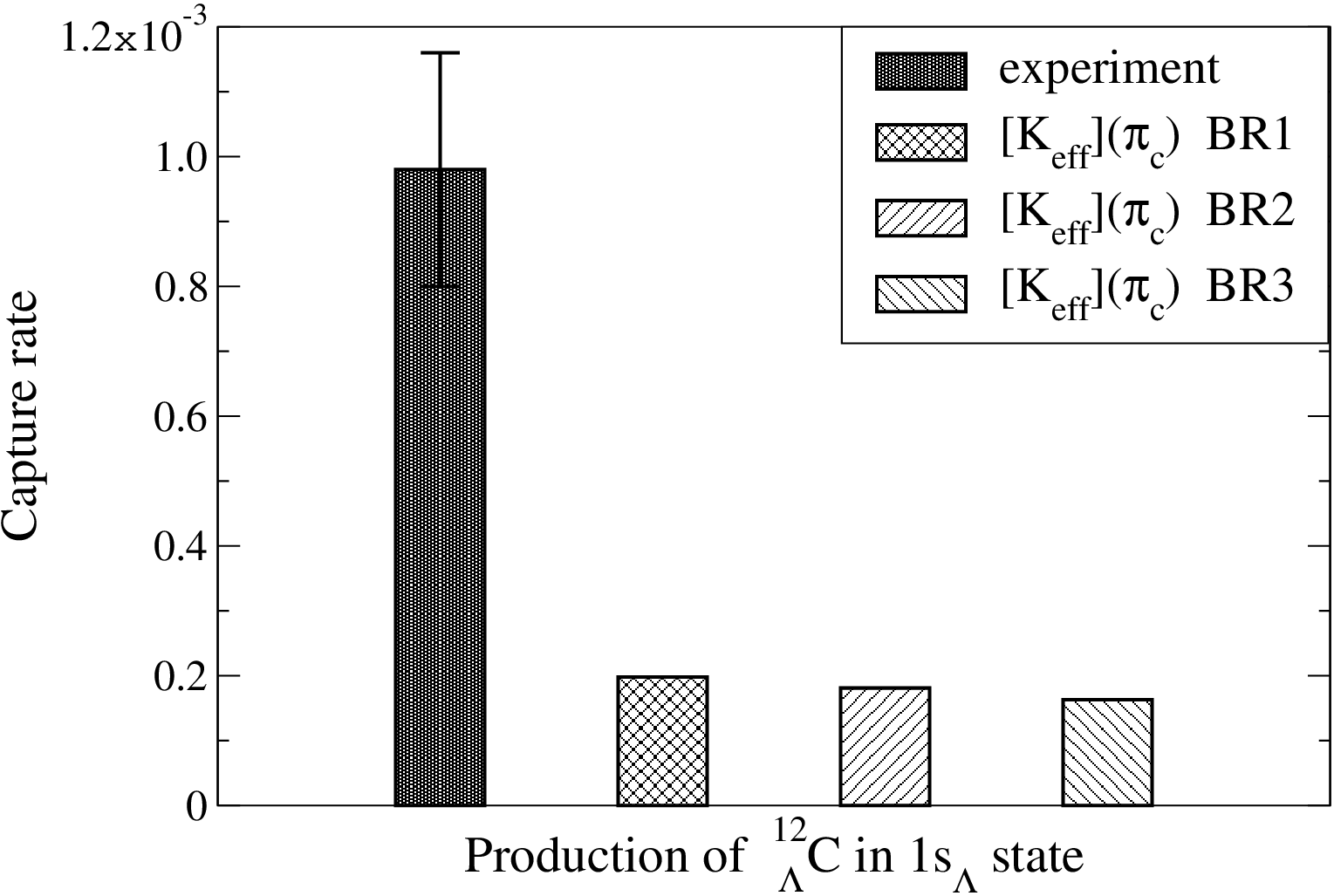} 
\caption{Sensitivity of calculated $1s_{\Lambda}$ capture rates per 
stopped $K^-$ in $^{12}{\rm C} \to {^{12}_{~\Lambda}{\rm C}}$ to  
$K^-n\rightarrow\pi^-\Lambda$ BRs, with respect to the measured summed 
$1s_{\Lambda}$ capture rate \cite{Tamura}.} 
\label{FigBRSens} 
\end{figure} 

We start by discussing the sensitivity of the calculated capture rates to 
the choice of BRs for the $K^-N\rightarrow\pi\Lambda$ capture 
process. In Fig.~\ref{FigBRSens}, we show the capture rate per stopped kaon 
Eq.~(\ref{eq:R/K}) calculated for the production of $^{12}_{~\Lambda}{\rm C}$ 
in the $1s_{\Lambda}$ state for the $K^-$-nucleus potential [$K_{\rm eff}$] 
and the pion-nucleus potential $\pi_c$. We recall that BR1 and BR2 come 
from our chiral microscopic model, whereas BR3 is derived from emulsion 
experiments. It is seen that all the calculated rates are quite close to each 
other but are significantly lower than the experimental data \cite{Tamura}. 
Since the difference between capture rates which correspond to various 
$K^-N\rightarrow\pi\Lambda$ BRs is relatively small (30\% at most) compared 
to uncertainties caused by other effects ($K^-$-nucleus or $\pi$-nucleus 
potential), in this paper, we present results based exclusively on BR1 
values which correspond to a well controlled in-medium chiral calculation.  

\begin{figure} 
\includegraphics[scale=0.4]{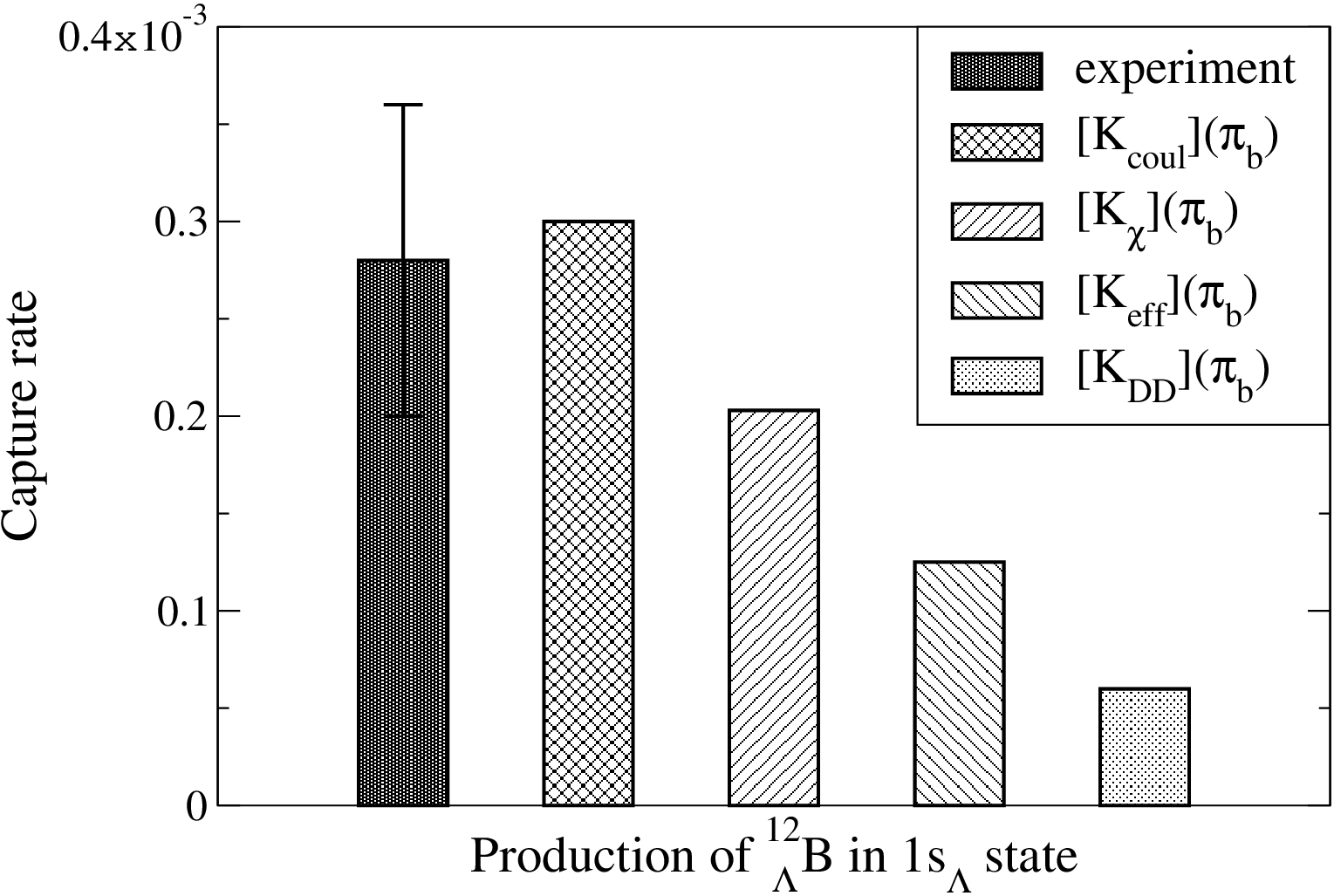} 
\caption{Sensitivity of calculated $1s_{\Lambda}$ capture rates per stopped 
$K^-$ in $^{12}{\rm C} \to {^{12}_{~\Lambda}{\rm B}}$ to $K^-$ wave functions, 
with respect to the measured summed $1s_{\Lambda}$ capture rate \cite{Ahmed}.} 
\label{FigKaonSens} 
\end{figure} 

The sensitivity to the $K^-$ wave functions for a given pion-nucleus potential 
($\pi_b$) is demonstrated in Fig.~\ref{FigKaonSens}, which shows calculated 
capture rates for the $1s_{\Lambda}$ state of $^{12}_{~\Lambda}{\rm B}$ 
with respect to the measured rate \cite{Ahmed} (figures for other targets 
look similar). The calculated rates are presented in order of increasing 
depth of the $K^-$-nucleus optical potential used to generate the kaon 
wave function, from a purely electromagnetic potential (zero depth) to the 
density-dependent potential [$K_{\rm DD}$] ($-190\,{\rm MeV}$ depth). 
The calculated capture rate appears to be a decreasing function of the 
$K^-$~-~nucleus potential depth. 

\subsubsection{Pion distorted waves} 

\begin{figure} 
\includegraphics[scale=0.4]{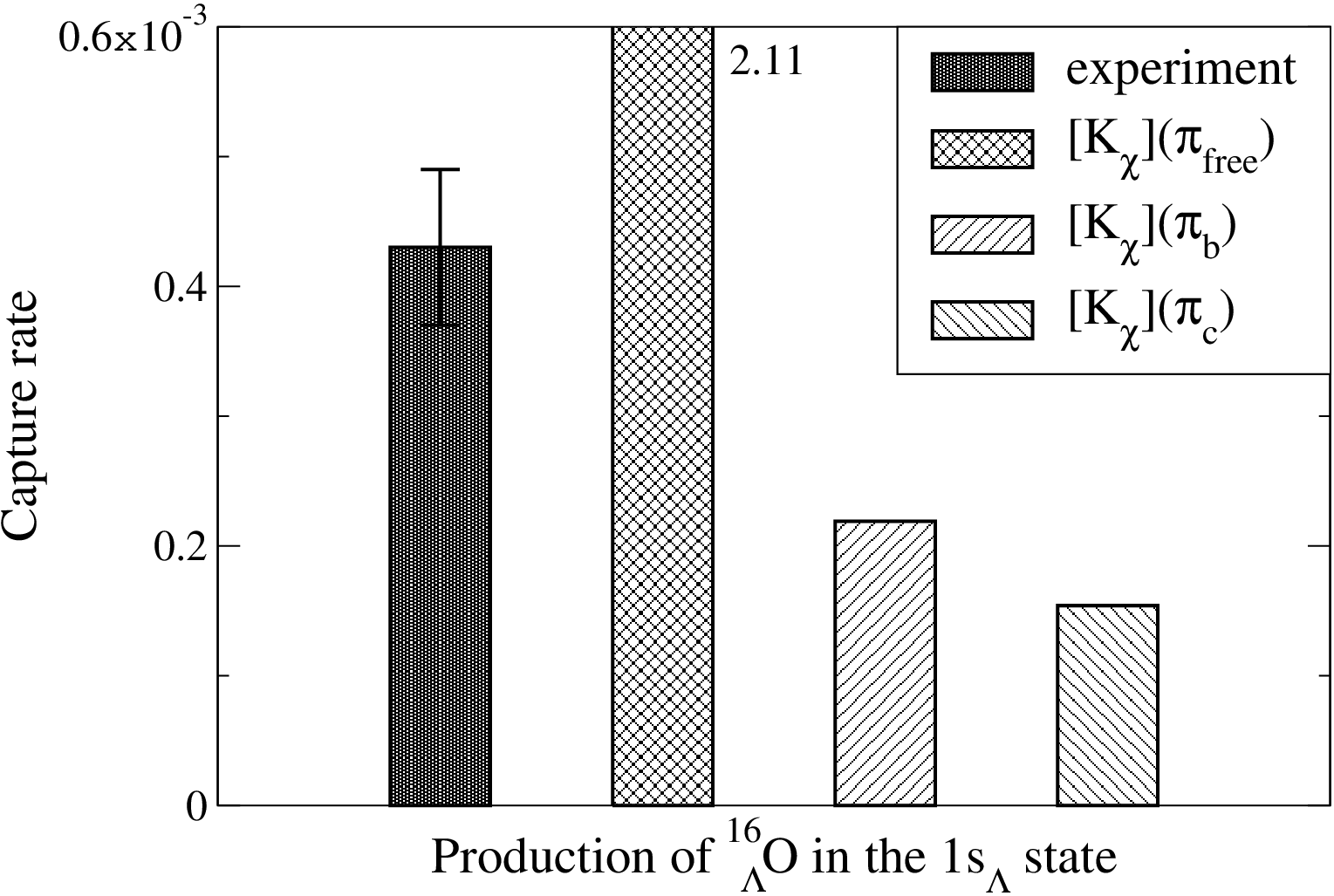} 
\caption{Sensitivity of calculated $1s_{\Lambda}$ capture rates per stopped 
$K^-$ in $^{16}{\rm O} \to {^{16}_{~\Lambda}{\rm O}}$ to pion distorted 
waves, with respect to the measured summed $1s_{\Lambda}$ capture rate 
\cite{Tamura}.} 
\label{FigPionSens} 
\end{figure} 

Figure~\ref{FigPionSens} shows the dependence of the capture rate 
on the choice of the pion-nucleus potential for the formation of 
$^{16}_{~\Lambda}{\rm O}$ in the $1s_{\Lambda}$ configuration, which consists 
of two separate peaks roughly of structure ($np^{-1}_{1/2}~{\Lambda} s_{1/2}$) 
and ($np^{-1}_{3/2}~{\Lambda} s_{1/2}$) for the $1^-$ g.s. and the 6 MeV first 
excited $1^-$ state \cite{Tamura}. One notes that pion distortion plays an 
important role. The difference between the results obtained with and without 
pion distortion is enormous. On the other hand, both pion optical potentials 
considered here lead to quite similar capture rates. The resulting rate for 
potential $\pi_c$ is just a little smaller than the result for potential 
$\pi_b$, with values $0.154\times 10^{-3}$ $(\pi_c)$ and $0.219\times 10^{-3}$ 
$(\pi_b)$, to be compared with a measured summed $1s_{\Lambda}$ capture rate 
of $0.43 \pm 0.06$ \cite{Tamura}.  

\subsection{Selected results} 

\subsubsection{$1s_{\Lambda}$ capture rates} 

\begin{table} 
\caption{Calculated capture rates per stopped $K^-$ (in units of $10^{-3}$) 
for the summed $1s_{\Lambda}$ production ($1^-$ transitions) in $p$-shell 
nuclei, using two variants for the $K^-$-atomic wave functions.} 
\label{TabCapRates1S} 
\begin{ruledtabular} 
\begin{tabular}{lrcccccl}
& $K^-$ potential & $^{7}_{\Lambda}{\rm Li}$ & $^{9}_{\Lambda}{\rm Be}$ & 
$^{12}_{~\Lambda}{\rm C}$ & $^{13}_{~\Lambda}{\rm C}$ & 
$^{16}_{~\Lambda}{\rm O}$ & \\ \hline 
& $[K_{\chi}]$ & 0.583 & 0.464 & 0.425 & 0.268 & 0.219 & \\  
& $[K_{\rm DD}]$ & 0.290 & 0.245 & 0.125 & 0.066 & 0.055 & \\ 
\end{tabular} 
\end{ruledtabular}
\end{table} 

The capture rates calculated for the summed production of $1s_{\Lambda}$ 
states in ($K^-_{\rm stop}, \pi^-$) reactions throughout the nuclear $p$ 
shell are assembled in Table~\ref{TabCapRates1S}. We only show results 
obtained with pion-nucleus potential $\pi_b$ and with kaon-nucleus 
potentials [$K_{\chi}$] and [$K_{\rm DD}$], which represent the two main 
directions for how the $K^-$-nucleus interaction is treated at present. 
It appears that the $1s_{\Lambda}$ capture rate is a decreasing function 
of $A$ throughout the nuclear $p$ shell, with a rate of decrease which depends 
sensitively on the depth of the $K^-$ nucleus potential. The ratio of 
$1s_{\Lambda}$ capture rate in $^{16}{\rm O}$ to that in $^{7}{\rm Li}$ is 
2.66 for [$K_{\chi}$] and 5.27 for [$K_{\rm DD}$]. Put differently, the ratio 
of rates related to the $K^-$ relatively shallow chiral potential to rates 
related to the $K^-$ relatively deep [$K_{\rm DD}$] potential increases 
throughout the $p$ shell, from approximately 2 for lithium up to about 4 for 
oxygen. This trend is caused by the node structure of the atomic wave 
functions used in the DWIA amplitudes Eq.~(\ref{eq:Il}). Whereas in 
$^7{\rm Li}$ the atomic $3D$ wave functions are nodeless within the nucleus, 
and the suppression of the rate for [$K_{\rm DD}$] with respect to 
[$K_{\chi}$] is caused by a node in (the real part of) the [$K_{\rm DD}$] $2P$ 
wave function, in $^{16}{\rm O}$ the [$K_{\rm DD}$] $3D$ wave function too has 
a node within the nucleus, which leads to further suppression with respect to 
the rate calculated for the [$K_{\chi}$] nodeless $3D$ wave function. Nodes of 
atomic wave functions within the nucleus are linked to the existence of 
quasibound $K^-$ {\it nuclear} states \cite{BFG97}. The deeper [$K_{\rm DD}$] 
potential generates such $L=1$ quasibound nuclear states throughout the $p$ 
shell and $L=2$ states beginning with the carbon isotopes, whereas the 
shallower [$K_{\chi}$] potential has $L=1$ states which only begin with the 
carbon isotopes and no $L=2$ quasibound nuclear states throughout the $p$ 
shell. 

We note the relatively sizable drop of the calculated $1s_{\Lambda}$ capture 
rates in going from $^{12}{\rm C}$ to $^{13}{\rm C}$. To identify its origin, 
we analyzed the impact of each one of the wave functions which appeare 
in the DWIA amplitude Eq.~(\ref{eq:Il}) in the carbon region (except for the 
remarkably stable $1s_{\Lambda}$ wave function) on the most important 
$|I^{\ell}_{1s,1p}|^2$ contribution. This procedure is demonstrated in 
Table~\ref{TabSensVse} where by starting with $^{12}{\rm C}$ related $K^-$, 
$\pi^-$, and neutron wave functions in the first row, we successively replaced 
them by $^{13}{\rm C}$ related wave functions as specified in the first column 
of rows $2-4$. The replacement of the $K^-$ wave function, in particular, 
is seen to have a small effect for the $[K_{\chi}]$ chiral potential 
wave function, but a large effect for the $[K_{\rm DD}]$ wave function. 
The successive replacement of other wave functions, particularly the 
oscillating pion distorted wave, leads to further suppression for both 
$[K_{\chi}]$ and $[K_{\rm DD}]$. 

\begin{table} 
\caption{Variation of the most dominant $|I^{\ell}_{1s,1p}|^2$ contribution, 
Eq.~(\ref{eq:Il}) for $L=2$  and pion distorted waves ($\pi_{\rm b}$) 
(in units of $10^{-12}$), upon going from $^{12}{\rm C}$ to $^{13}{\rm C}$.} 
\label{TabSensVse} 
\begin{ruledtabular} 
\begin{tabular}{lrccl} 
& path of variation & $|I^3_{1s,1p}|^2~({[K_{\chi}]})$ & 
$|I^1_{1s,1p}|^2~({[K_{\rm DD}]})$ & \\ \hline 
& $^{12}{\rm C} $	& 3.92 & 1.44 & \\ 
& + kaon w.f. & 3.90 & 0.74 & \\ 
& + pion w.f. & 2.77 & 0.51 & \\ 
& + neutron w.f. = $^{13}{\rm C}$ & 2.70 & 0.42 & \\ 
\end{tabular} 
\end{ruledtabular} 
\end{table} 

In Table~\ref{TabSensRho}, we show the variation of the effective nuclear 
density Eq.~(\ref{efektivnihustota}), which appears in the denominator 
of Eq.~(\ref{eq:R/Y}) for the capture rate. The resultant effect is 
considerably more moderate than for the $[K_{\rm DD}]$ DWIA amplitude listed 
in Table~\ref{TabSensVse}, simply because the two components of the 
${\bar\rho}_N$ integrand are each positive definite and are less sensitive 
to the position of nodes in the $[K_{\rm DD}]$ $3D$ atomic wave function. 
The net outcome of the variations studied in Tables \ref{TabSensVse} and 
\ref{TabSensRho} is a considerably smaller capture rate for $^{13}{\rm C}$ 
than for $^{12}{\rm C}$, at least for $[K_{\rm DD}]$. For $[K_{\chi}]$, 
the reduction appears weaker (it is more effective in some of the other 
$|I^{\ell}_{1s,1p}|^2$ contributions not shown here). 

\begin{table} 
\caption{Variation of ${\bar\rho}_N$ Eq.~(\ref{efektivnihustota}) for $L=2$ 
(in units of $10^{-10}$) upon varying the $^{12}{\rm C}$ and $^{13}{\rm C}$ 
input.} 
\label{TabSensRho} 
\begin{ruledtabular} 
\begin{tabular}{lrccccl} 
& $K^-$(C)&\multicolumn{2}{c}{[$K_{\chi}$]}&\multicolumn{2}{c}{[$K_{\rm DD}$]} 
& \\
&& $\rho_N(^{12}{\rm C})$ & $\rho_N(^{13}{\rm C})$ & $\rho_N(^{12}{\rm C})$ & 
$\rho_N(^{13}{\rm C})$ & \\ \hline 
& $K^-(^{12}{\rm C})$  & 2.47 & 2.88 & 3.41 & 3.98 & \\ 
& $K^-(^{13}{\rm C})$  & 2.55 & 2.97 & 2.80 & 3.27 & \\ 
\end{tabular} 
\end{ruledtabular} 
\end{table} 

\begin{table} 
\caption{Calculated capture rates per stopped $K^-$ (in units of $10^{-3}$) 
for production of $1s_{\Lambda}$ states ($1^-$ transition) and $1p_{\Lambda}$ 
states ($0^+$ and $2^+$ transitions) and selected experimental rates.} 
\label{TabCapRates1P} 
\begin{ruledtabular} 
\begin{tabular}{lcccccl} 
& transition & input & $^{12}_{~\Lambda}{\rm B}$ \cite{Ahmed} & 
$^{12}_{~\Lambda}{\rm C}$ \cite{Tamura}& 
$^{16}_{~\Lambda}{\rm O}$ \cite{Tamura} & \\  \hline  
& $1^-$ & [$K_{\chi}$]  & 0.203 & 0.425 & 0.219 &    \\
&      & [$K_{\rm DD}$] & 0.060 & 0.125 & 0.055 & \\   
&      & exp. rates & $0.28\pm 0.08$ & $0.98\pm 0.12$ & $0.43\pm 0.06$ & \\ 
& $0^+$ & [$K_{\chi}$]  & 0.096 & 0.216 & 0.134 & \\
&      & [$K_{\rm DD}$] & 0.011 & 0.021 & 0.020 & \\   
& $2^+$ & [$K_{\chi}$]  & 0.547 & 1.052 & 0.872 & \\ 
&      & [$K_{\rm DD}$]  & 0.192 & 0.410 & 0.330 & \\   
& $0^+~+~2^+$ & [$K_{\chi}$]  & 0.643 & 1.268 & 1.006 & \\ 
&            & [$K_{\rm DD}$] & 0.203 & 0.431 & 0.350 & \\ 
&       & exp. rates & $0.35 \pm 0.09$ & $2.3 \pm 0.3$ & $1.68 \pm 0.16$ & \\ 
\end{tabular} 
\end{ruledtabular} 
\end{table} 

\subsubsection{$1p_{\Lambda}$ capture rates} 

Calculated capture rates for the $1p_N \rightarrow 1p_{\Lambda}$ $0^+$ 
and $2^+$ transitions [see Eq.~(\ref{eq:R/Y})] are presented in 
Table~\ref{TabCapRates1P} for $^{12}{\rm C}$ and $^{16}{\rm O}$ targets. 
For completeness, we also included the $1p_N \rightarrow 1s_{\Lambda}$ $1^-$ 
transition discussed in Sec.~IV B1 and added experimental rates from 
KEK \cite{Tamura} and BNL \cite{Ahmed}. Here, the reason for choosing 
KEK over CERN \cite{Faessler} and FINUDA \cite{Agnello} is that for 
$^{12}_{~\Lambda}{\rm C}$ production, the KEK \cite{Tamura} rates are the 
closest to the isospin factor 2 expected in getting these rates from the 
$^{12}_{~\Lambda}{\rm B}$ rates which were measured only at 
BNL \cite{Ahmed}.{\footnote{The reported summed $1s_{\Lambda}$ capture rate 
in $^{12}_{~\Lambda}{\rm C}$ varies from $(0.2 \pm 0.1) \times 10^{-3}/K^-$ 
\cite{Faessler} to $(1.86 \pm 0.14) \times 10^{-3}/K^-$ \cite{Agnello}.}} 
Of course, if further ($K^-_{\rm stop}, \pi^0$) experiments are done on 
$^{12}{\rm C}$, which would lead to different results from those of 
Ref.~\cite{Ahmed}, one's preference might change accordingly.{\footnote
{The ratio of the calculated $1s_{\Lambda}$ capture rate to 
$^{12}_{~\Lambda}{\rm C}$ over that of $^{12}_{~\Lambda}{\rm B}$ is largely 
caused by the ratio of the $K^- N\to\pi\Lambda$ BRs which in the limit of 
good isospin is 2. A slight departure from this ratio is caused by 
charge-dependent effects in our calculation, notably from the outgoing pion 
DWs.}} Similar to the summed $1s_{\Lambda}$ rate discussed in Sec.~IV B1, the 
summed $1p_{\Lambda}$ rate for the deep potential [$K_{\rm DD}$] is smaller 
by a factor $3-4$ than for the relatively shallow potential [$K_{\chi}$]. 
Excluding the old CERN data \cite{Faessler}, the calculated $1p_{\Lambda}$ 
capture rates are generally smaller than the experimentally reported rates, 
with the exception that the [$K_{\chi}$] rates for $^{12}_{~\Lambda}{\rm B}$ 
are larger than the reported rate \cite{Ahmed}. Since the $1p_{\Lambda}$ 
spectral strength is partly mixed into the ($K^-_{\rm stop}, \pi$) quasifree 
continuum, and its extraction from the measured spectra is considerably more 
ambiguous than the extraction of the summed $1s_{\Lambda}$ production rate, 
we do not proceed further to confront theory with experiment for the summed 
$1p_{\Lambda}$ production rate.

\section{Conclusion} 

We performed DWIA ($K^-_{\rm stop}, \pi$) calculations for $p$-shell targets, 
using several $K^-$ and pion wave functions to test the sensitivity of the 
calculated hypernuclear capture rates to the choice of these wave functions. 
The calculated capure rates are generally smaller than the measured ones; 
the deeper the $K^-$ potential, the smaller is the capture rate. Since the 
absolute normalization of capture at rest experimental rates is a delicate 
matter, we suggest to focus on the $A$ dependence of the measured rates, 
expecting it to be largely free of the absolute normalization of the data.  
The calculated capture rates for a given $K^-$ optical potential decrease as 
a function of $A$, with the fractional difference between the rates calculated 
for the two extreme $K^-$ optical potentials, the shallow [$K_{\chi}$] and the 
deep [$K_{\rm DD}$], increasing steadily with $A$. We find other dependencies 
of the calculated capture rates to be secondary to the dependence on the 
$K^-$-atomic wave function in the range studied here. We argue that 
a dedicated experimental study of $1s_{\Lambda}$ capture rates in $p$-shell 
targets, such as reported recently by the FINUDA Collaboration in 
a preliminary form \cite{BonomiHYPX}, could yield useful information on the 
depth of the threshold $K^-$ optical potential by comparing the measured $A$ 
dependence with the $A$ dependence of the calculated capture rates listed in 
Table~\ref{TabCapRates1S}.

\section*{Acknowledgments}

This work was supported by the GAUK Grant No. 91509 and the GACR Grant 
No. 202/09/1441, as well as by the EU initiative FP7, HadronPhysics2, 
under Project No. 227431.

%\newpage 


\begin{thebibliography}{99} 

\bibitem{Faessler} M.A. Faessler {\it et al.}, Phys. Lett. B {\bf 46}, 468 
(1973). 

\bibitem{Tamura} H. Tamura, R.S. Hayano, H. Outa, and T. Yamazaki, 
Prog. Theor. Phys. Suppl. {\bf 117}, 1 (1994). 

\bibitem{Ahmed} M.W. Ahmed {\it et al.}, Phys. Rev. C {\bf 68}, 064004 (2003). 

\bibitem{Agnello} M. Agnello {\it et al.}, Phys. Lett. B {\bf 622}, 35 (2005). 

\bibitem{HYP06} M. Agnello {\it et al.}, in {\it Hypernuclear and Strange 
Particle Physics HYP2006}, edited by J.~Pochodzalla and Th.~Walcher 
(SIF and Springer, Berlin Heidelberg, 2007) p. 57. 

\bibitem{BonomiHYPX} M. Agnello {\it et al.}, Nucl. Phys. A {\bf 835}, 414 
(2010) [presented by G. Bonomi for the FINUDA Collaboration at HYP-X, Tokai, 
Sept. 2009]. 

\bibitem{HLW74} J. H\"{u}fner, S.Y. Lee, and H.A. Weidenm\"{u}ller, 
Nucl. Phys. A {\bf 234}, 429 (1974). 

\bibitem{GalKlieb} A. Gal and L. Klieb, Phys. Rev. C {\bf 34}, 956 (1986). 

\bibitem{MatsuyamaYazaki} A. Matsuyama and K. Yazaki, Nucl. Phys. A {\bf 477}, 
673 (1988). 

\bibitem{CieplyFriedman} A. Ciepl\'{y}, E. Friedman, A. Gal, and J. Mare\v{s}, 
Nucl. Phys. A {\bf 696}, 173 (2001). 

\bibitem{KC10} V.~Krej\v{c}i\v{r}\'{i}k and A.~Ciepl\'{y}, Acta Phys. Pol. B 
{\bf 41}, 317 (2010). 

\bibitem{KaiserSiegelWeise} N. Kaiser, P.B. Siegel, and W. Weise, 
Nucl. Phys. A {\bf 594}, 325 (1995). 

\bibitem{WaasKaiserWeise} T. Waas, N. Kaiser, and W. Weise, Phys. Lett. B 
{\bf 365}, 12 (1996). 

\bibitem{OsetRamos} E. Oset and A. Ramos, Nucl. Phys. A {\bf 635}, 99 (1998). 

\bibitem{BorasoyNisslerWeise} B. Borasoy, R. Nissler, and W. Weise, 
Eur. Phys. J. A {\bf 25}, 79 (2005). 

\bibitem{CieplySmejkal} A. Ciepl\'{y} and J. Smejkal, Eur. Phys. J. A 
{\bf 34}, 237 (2007). 

\bibitem{CieplySmejkalPrep} A. Ciepl\'{y} and J. Smejkal, Eur. Phys. J. A 
{\bf 43}, 191 (2010). 

\bibitem{Lutz} M. Lutz, Phys. Lett. B {\bf 426}, 12 (1998). 

\bibitem{VanDerVelde} C. Vander Velde-Wilquet {\it et al.}, Nuovo Cimento A 
{\bf 39}, 538 (1977). 

\bibitem{Batty} E. Friedman, A. Gal, and C.J. Batty, Nucl. Phys. A {\bf 579}, 
518 (1994). 

\bibitem{BattyComunication} C.J. Batty (private communication, 1995). 

\bibitem{HashimotoTamura} O. Hashimoto and H. Tamura, Progr. Part. Nucl. Phys. 
{\bf 57}, 564 (2006). 

\bibitem{EricsonEricson} M. Ericson and T.E.O. Ericson, Ann. Phys. {\bf 36}, 
323 (1966). 

\bibitem{pionoptpot1} H.A. Thiessen {\it et al.}, LAMPF Report No. LA-7607-PR, 
1978 (unpublished). 

\bibitem{pionoptpot2} C.J. Harvey {\it et al,}, LAMPF Report No. 
LA-UR-84-1732, 1984 (unpublished). 

\bibitem{BFG97} C.J. Batty, E. Friedman, and A. Gal, Phys. Rep. {\bf 287}, 
385 (1997). 

\end{thebibliography}
\end{document}